\documentclass[%
 preprint,%
 aps,
prl,superscriptaddress,showkeys
]{revtex4-2}

\usepackage{graphicx}
\usepackage{dcolumn}
\usepackage{bm}
\usepackage{amsmath}
\usepackage{setspace}
\usepackage{placeins}
\usepackage{color}
\usepackage{ulem}

\begin{document}
\doublespacing
\preprint{}

\title{Place-cell heterogeneity underlies power-laws in hippocampal activity}
\date{July 28, 2025}

\author{J.J. Briguglio}
\affiliation{Janelia Research Campus, 19700 Helix Drive, Ashburn, VA 20147, USA}
\author{J. Lee}
\affiliation{Janelia Research Campus, 19700 Helix Drive, Ashburn, VA 20147, USA}
\affiliation{Howard Hughes Medical Institute, Beth Israel Deaconess Medical Center, 330 Brookline Ave., Boston, MA 02215, USA.}
\author{A.K. Lee}
\affiliation{Janelia Research Campus, 19700 Helix Drive, Ashburn, VA 20147, USA}
\affiliation{Howard Hughes Medical Institute, Beth Israel Deaconess Medical Center, 330 Brookline Ave., Boston, MA 02215, USA.}
\author{V. Hakim}
\email{Vincent.Hakim@phys.ens.fr}
\affiliation{
Laboratoire de Physique de l'École Normale Supérieure, ENS, Université PSL,
CNRS, Sorbonne Université, Université Paris Cité, Paris, France}
\author{S. Romani}
\email{RomaniS@janelia.hhmi.org}
\affiliation{Janelia Research Campus, 19700 Helix Drive, Ashburn, VA 20147, USA}



\begin{abstract}
Power-law scaling in coarse-grained data suggests critical dynamics, but the true source of this scaling often remains unclear. Here, we analyze neural activity recorded during spatial navigation, reproducing power-law scaling under a phenomenological renormalization group (PRG) procedure that clusters units by activity similarity. Such scaling was previously linked to criticality. Here, we show that the iterative nature of the procedure itself leads to the emergence of power laws  when applied to heterogeneous, non-interacting units obeying spatially structured activity without requiring critical interactions. Furthermore, the scaling exponents produced by heteregeneous non-interacting units match the observed exponents in recorded neural data. A simplified version of the PRG further reveals how heterogeneity smooths transitions across scales, mimicking critical behavior. The resulting exponents depend systematically on system and population size, predictions confirmed by subsampling the data.
\end{abstract}

\maketitle


\section{Introduction}
Renormalization group (RG) techniques probe interacting systems for power-law scaling, revealing critical phenomena \cite{Kadanoff1966, Wilson71, WilsonKogut1974,Amit2005}. Inspired by RG, coarse-graining methods applied to empirical data can extract critical exponents to classify system dynamics \cite{Meshulam2018Arxiv,Meshulam2019,Berman2023,Garuccio2023,Sooter2024}, yet power laws can also stem from non-critical processes, obscuring interpretation when system details are limited \cite{Schwab2014, Aitchison2016, Touboul2017a, Stringer2019, Faqeeh2019, Cubero2019, Morrell2021,Priesemann2018}. This ambiguity complicates analyzing empirical datasets, including those from biological systems, where coarse-graining often suggests critical states \cite{Meshulam2019,Ponce2023,Morales2023}.

Recent studies applying a phenomenological renormalization group (PRG) to neural activity from mice navigating spatially have reported power-law scaling, suggesting criticality \cite{Meshulam2018Arxiv,Meshulam2019}. Unlike traditional models of spatially tuned units that fail to replicate this scaling \cite{Meshulam2018Arxiv, Morrell2021}, these findings fueled debate over its mechanisms \cite{Nicoletti2020,Touboul2017a,Morales2023,Cubero2019}. We approach this using data from a large spatial environment (40 m), where hippocampal CA1 neurons exhibit heterogeneous place field distributions that can be modeled as independent Poisson processes with Gamma-distributed rates \cite{Rich2014, Lee2020}. After reproducing these previous findings, we demonstrate that this spatial heterogeneity alone generates the observed scaling under PRG, with non-interacting units producing power laws that mimic signatures of criticality. This reveals that PRG can transform unit-level variability into apparent scale invariance, reshaping power-law interpretations across complex systems.

\section{Results}
Transgenic mice (n=4) expressing GCaMP6f were trained to run on a spherical treadmill (40cm diameter) in multiple 40m-long virtual environments (Figure 1A). Water rewards were provided randomly throughout the environment and the animals were ‘teleported’ to the start after running the entire length of the track. Calcium activity of dorsal hippocampal CA1 pyramidal neurons was imaged, yielding 700-870 neurons per session (Figure 1B). In each session, the animals ran 8 laps in a single familiar environment. Calcium signals were filtered and normalized to yield binary signals representing when a cell was active (Figure 1C). Additional experimental and data preprocessing details are presented in \cite{Lee2020}.
\begin{figure*}[t]
\centering
\includegraphics[width=.85\textwidth]{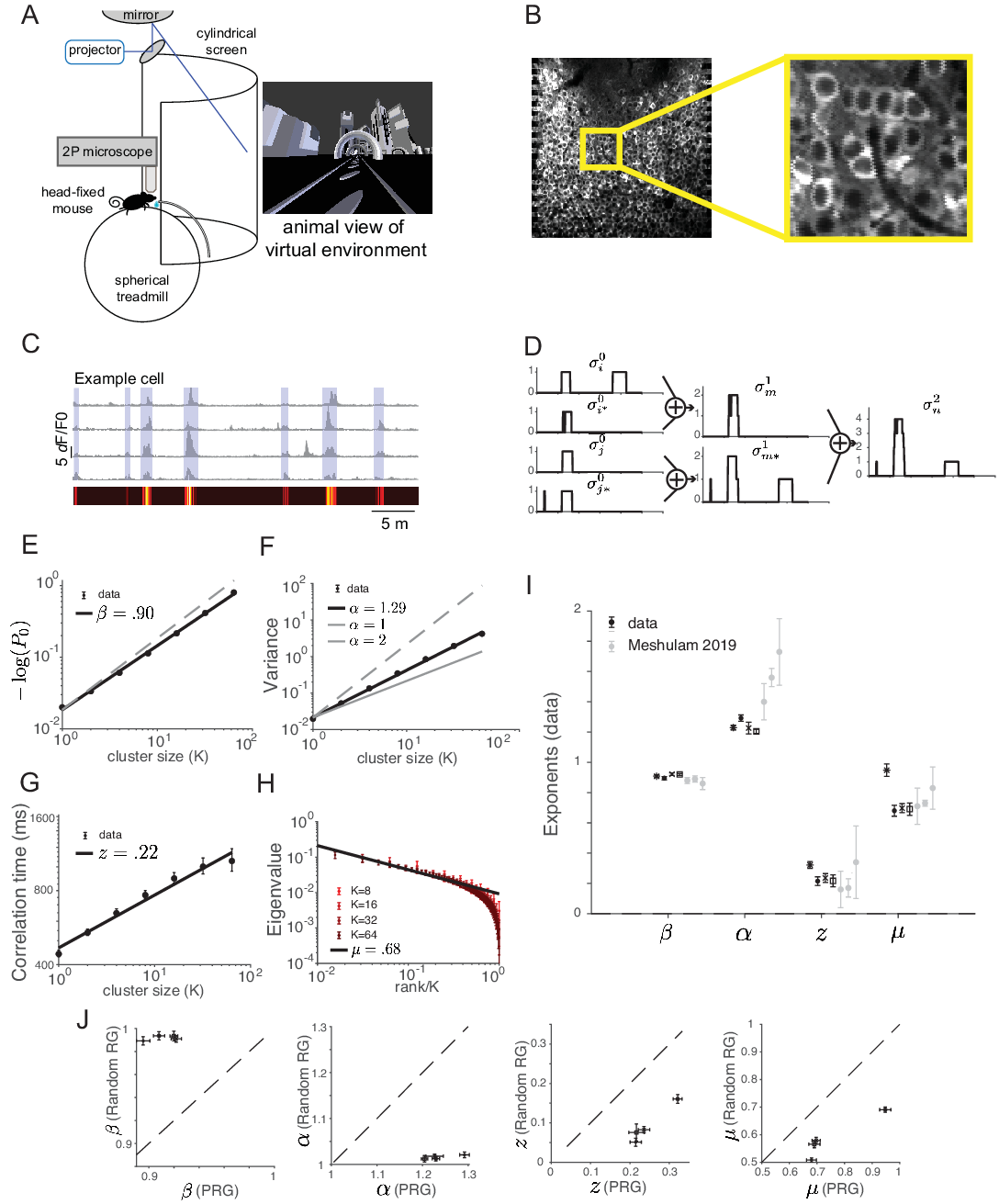}
\caption{Figure 1: Experimental schematic. A) A transgenic mouse runs on a spherical treadmill that moves the animal through a virtual environment projected onto a surrounding screen. B) Hippocampal cells are viewed via calcium-imaging, giving hundreds of cells per recording. C) An example cell exhibits peaks in calcium response consistently across laps in multiple locations. D) Phenomenological renormalization group procedure is performed on binarized versions of the single-cell signals by combining cells with most correlated activity in a recursive fashion. This produces power law scaling for E) log-probability of silence, F) within-cluster eigenvalue spectrum, G) correlation time, H) variance. Power-law scaling exponents across mice is summarized in I), with each marker representing a different mouse. J) Comparison of measured exponents under maximal-pair combining (PRG) and random-pair combining (Random RG).}
\end{figure*}

The binarized activity traces from $N$ recorded neurons, $\sigma_i^0(t)$,$i=1..N$ were used as the inputs to the Phenomenological Renormalization Group (PRG) analysis used in \cite{Meshulam2019}. Briefly, coarse-graining was achieved by choosing the pair of neurons with the maximal correlation between their activity traces,  removing this pair, and iterating until no more pairs can be formed. The activity of each coarse-grained unit after this first iteration, $\sigma_i^1(t)$,$i=1..\lfloor N/2\rfloor$, was defined to be the sum of the activity of the constituent pair of neurons.
This process was repeated as long as there remained at least $N_{min}=4$ units, summing the activity traces of all paired units at each stage (Figure 1D). After $k$ iterations, the activity of each of the $N_k=\lfloor N/2^k \rfloor$ units $\sigma_i^k (t)$, was a sum of the activity of a cluster of $K=2^k$ recorded neurons. At each step, the probability of silence ($P_0(k)=\langle P(\sigma_i^k(t)=0) \rangle_{N_k,t}$), the mean variance of individual units ($\mathrm{Var}(K)=\langle \langle\sigma_i^k(t)^2\rangle_t -\langle\sigma_i^k(t)\rangle_t^2 \rangle_{N_k}$), the correlation time ($\tau_{corr}(K)$ exponential timescale fit to the average autocorrelation function $\langle R_{\sigma_i^k\sigma_i^k}(\Delta t)\rangle_{N_k}$), and the within-cluster eigenvalue spectra are computed. The within-cluster eigenvalue spectra is $\lambda_r(K)=\langle \mathrm{Eigv}(C(m,k),r) \rangle_{N_k}$ where $\mathrm{Eigv}(X,r)$ returns the $r^{th}$ largest eigenvalue of matrix $X$. The matrix $C(m,k)$ is the covariance matrix of the units that comprise the m-th coarse-grained unit at the k-th PRG iteration i.e the $\sigma_{i_m}^0$ with $\sum_{i_m} \sigma^0_{i_m}=\sigma^k_m$. Three exponents are computed by fitting a first-order polynomial to the log-transformed relationships described by $\mathrm{Var}\propto K^\alpha$, $-\log(P_0)\propto K^\beta $, and $\tau_{corr}\propto K^z$. The within-cluster eigenvalue spectrum was fit to the relationship $\lambda_r=A(K/r)^\mu$ where $\lambda_r$ denotes the $r^{th}$ largest eigenvalue, see \cite{Meshulam2018Arxiv,Meshulam2019}. We used a first-order polynomial fit to the corresponding log-transformed relationship on the four largest $K$ values per session excluding $r/K>.5$ in order to capture the scaling of the larger eigenvalues, rather than the bulk of low eigenvalues, as expected for these matrices, see e.g. \cite{Hu2022}.

We sought to reproduce the power-law scaling observed following the application of the PRG procedure to hippocampal CA1 data. Silence probability, variance, correlation time, and eigenvalue spectrum were each well described by power-laws (Figure 1E-H). We noted consistency in the power-law exponents across mice (Figure 1I, black markers). This analysis qualitatively reproduces previous findings \cite{Meshulam2018Arxiv,Meshulam2019}, though we observed systematic differences in the values of some of the previously reported exponents, most noticeably in the exponent associated to the scaling of the variance (Figure 1I, compare black with grey markers). 

\begin{figure*}[t]
\centering
\includegraphics[width=\textwidth]{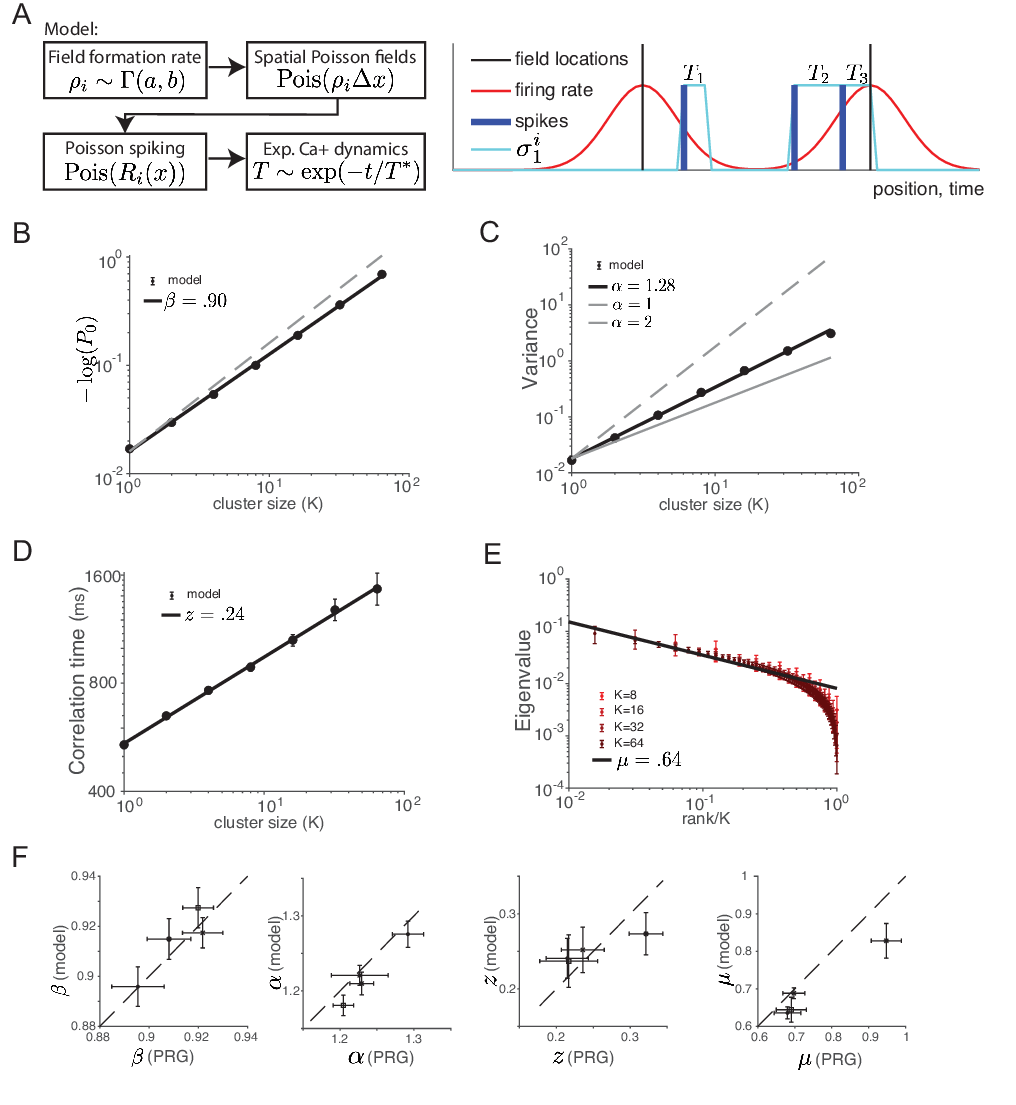}
\caption{Figure 2: Hippocampal spatial coding model recapitulates measured exponents. A) Schematic of spatial coding model using parameters derived from experimental data. Power-law scaling is observed for B) log-probability of silence, C) variance, D) autocorrelation time, and E) within-cluster eigenvalue spectrum. F) Comparison between exponents measured from hippocampal spatial coding model and directly from the data across animals.}
\end{figure*}

To establish the significance of these results, we first built a simple control coarse-graining procedure that pairs neurons randomly (Random PRG), rather than based on their activity correlation. Using this procedure, we observed power-law scaling behavior with trivial scaling exponents for $\alpha\approx1$ and $\beta\approx1$, as expected from combining uncorrelated Gaussian processes. These control exponents are far from those resulting from the PRG procedure (Figure 1J). We similarly observed different values for $z$ and $\mu$ relative to our control. This demonstrates that the correlation structure in the data gives rise to non-trivial exponents (Fig 1).

We next considered the possible contribution of spatially tuned firing, a hallmark of hippocampal activity, to the power laws observed with the application of the PRG procedure. We used an accurate statistical model of CA1 place field activity, composed of model neurons with independent (conditionally on spatial location), heterogeneous, reliable place fields (Figure 2A) \cite{Rich2014, Lee2020}. The probability of neuron $i$ having a field in some region was defined as a Poisson process, $P_i (x)=1-e^{(-\rho_i \Delta x)}$, where $x$ denotes the spatial bin of size $\Delta x=1 cm$. The Poisson rate $\rho_i$ was heterogeneous across cells, following a gamma distribution $G(\rho;a,b)=b^a\rho^{a-1}e^{-b\rho}/\Gamma(a)$ with shape parameter $a$ and rate parameter $b$, both fit to match the measured field numbers. Each simulated neuron had a modeled calcium spike rate described by $R_i (x(t))=B+A\sum_k \exp\left[-(x(t)-x_k^{(i)} )^2/2s^2\right]$ where $B$ is a background transient rate, $A$ is the peak amplitude of the place fields, $x_k^{(i)}$ denotes the field locations for unit $i$,$\sum_k$ sums over all place fields, and $s=10 cm$ determines the field size. The field amplitude $A$ and the background rate $B$ were fit to match the measured population transient numbers. We simulated an animal running through the environment at a constant speed ($v=30 cm/s$), $x(t)=vt$  and generated calcium spikes (sampled at 30 Hz) as a Poisson process with rate based on the animal’s location at each time. To mimic calcium transient durations measured from our imaging recordings, each calcium spike was replaced by a signal of constant amplitude for a duration $T$, sampled from the distribution $P(T)={T^*}^{(-1)} e^{(-T /T^* )}$, where $T^*$ was the average transient duration estimated from the recordings, separately for each session (Figure 2A). The fit parameters are presented in Table~\ref{tab:parameters}.

\begin{figure*}[t]
\centering
\includegraphics[width=\textwidth]{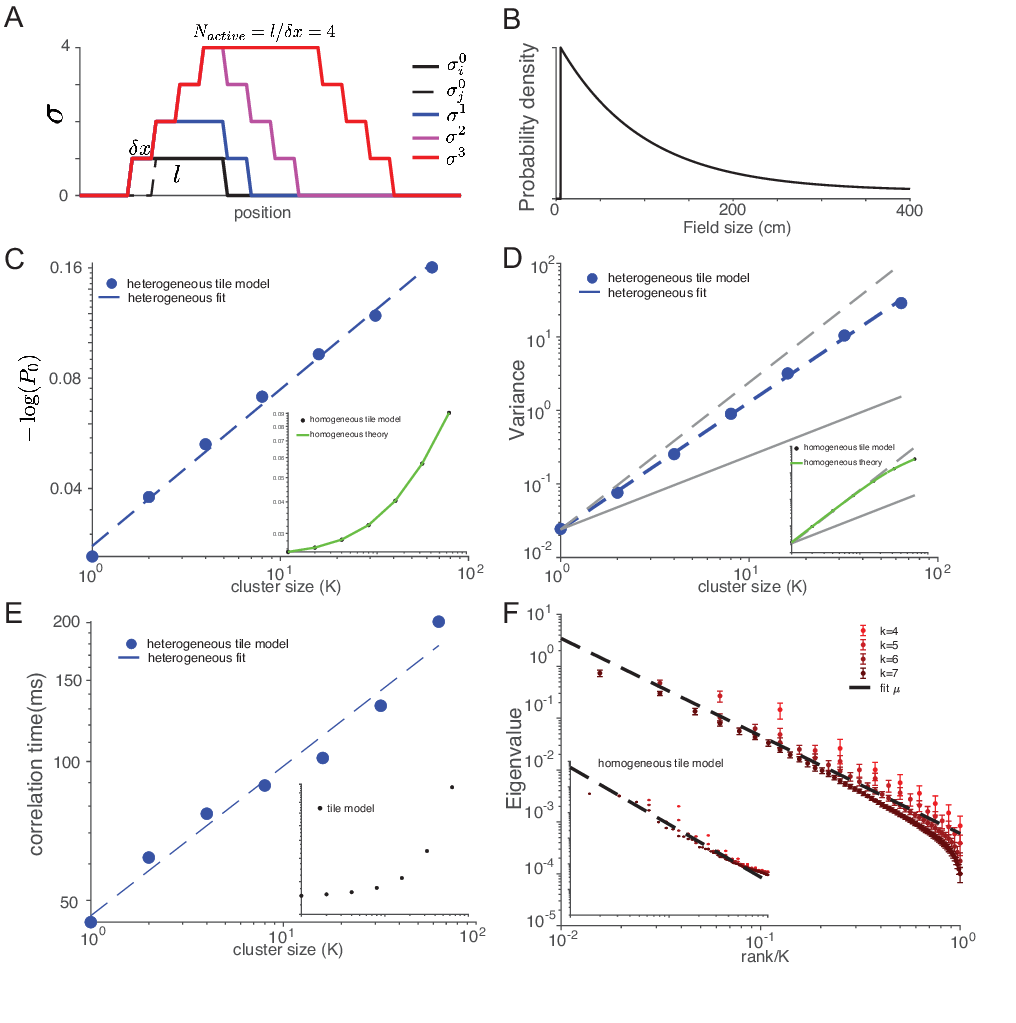}
\caption{Figure 3: Simplified spatial coding model. A) Individual unit responses as coarse-graining progresses show a distinct change in character around $K=\rho$. Significant heterogeneities in spatial response size (B) leads to apparent power law scaling not visible to homogeneous model (insets) for C) log-probability of silence, D) variance, E) autocorrelation time, and F) within-cluster eigenvalue spectrum.}
\end{figure*}

The simulated data was then analyzed in the same manner as the measured empirical data. We observed power-law scaling for the silence probability (Figure 2B), the variance (Figure 2C), the correlation time (Figure 2D), and the eigenvalue spectra (Figure 2E). Additionally, the corresponding exponents are consistent with those derived from the empirical data (Figure 2F). Hence, a model with independent neurons, whose correlations solely arise from spatial coding, could reproduce the power-law scaling exponents extracted from the data.

We next sought to understand how this model could give rise to power laws. We  constructed an approximate PRG procedure based on a simplified model of spatial coding neurons consisting of a population of binary place cells whose fields tile the environment.
We consider first a model with homogeneous fields that are binary with size $l$ and that uniformly tile the environment (field center of unit $i$ is $x_i=Li/N$ with $L$ total environment size and $N$ the population size). The activity of unit $i$ without any coarse-graining (iteration $k=0$) is $\sigma_i^0 (x)=\Theta(l/2-|x-x_i|)$ where $\Theta$ is the Heaviside function ($1$ when the argument is positive, $0$ otherwise) and $x$ is the position of the simulated animal. In this model units with adjacent place fields are the most correlated. Therefore, the coarse graining procedure is simply defined as summing the activity of pairs of neighboring units at every iteration (Figure 3A). As shown below, this homogeneous tile model does not exhibit power-law scaling with coarse-graining, but it provides useful insights into the properties of a system where correlations are solely due to spatial tuning of the individual neurons. Note that the approximate PRG is similar in spirit to simple models that were previously proposed to describe the scaling properties of coarsening breadth figures \cite{Derrida1991}.
 
We simulated this homogeneous tile model using parameters approximating the reference data (environment size ~40m, field size ~1m , population size ~1000) with an animal running at constant speed (30 cm/s) and used the PRG method described above. We observe that the probability of silence (Fig 3B inset), variance (Fig 3C inset), correlation time (Fig 3D inset), and eigenvalue spectrum (Fig 3E inset) do not follow an exact power-law scaling. Similar findings were observed to the exact coarse graining described (shown) and using PRG with random permutations to break correlation symmetry (not shown). A mathematical analysis of this model (see End Matter) shows the presence of non-uniform scaling properties, determined by the number of units active at any given location, $N_{active}=l/\delta x$, where $\delta x=L/N$ is the spacing between neighboring units ($N_{active}=25.6$ for the chosen parameters). A transition in scaling properties occurs at cluster size $K^*=N_{active}$, where coarse-grained units’ field amplitudes continue to increase for $K<N_{active}$ and only field width increases for $K>N_{active}$. For instance, the variance, $Var(\sigma^K)=\sum_{i=0}^3a_i(K) K^i$, has different coefficients $a_i(K)$ for $K>N_{active}$ and $K<N_{active}$. Our analysis perfectly predicts the scaling properties for the probability of silence and variance (Figure 3C,D insets). We did not derive a closed-form expression for the correlation time and eigenvalue spectrum, but the simulated results do not exhibit power-law scaling (Fig 3E,F insets). 

The most salient difference between the simple homogeneous tile model described above and our hippocampal place-field simulation (Fig.~2), and previous studies \cite{Meshulam2018Arxiv,Morrell2021}, is the presence of heterogeneity in spatial coding across neurons, which can be best quantified in large environments \cite{Lee2020}. We explored the effect of such an heterogeneity by considering a heterogeneous tile model in which the fields are drawn according
to a Gamma-Poisson distribution of field sizes (field size $l_0+l_r G(a_r,b_r)$ with $l_0=5cm$, $l_r=95cm$, $a_r=1$, and $b_r=1$). These values of $a_r$ and $b_r$ are similar to those observed in mice recordings and are chosen for simplicity since the Gamma-Poisson distribution simply reduces to an exponential distribution. We observed scaling more closely matched to a power-law for the probability of silence (Fig 3B), variances (Fig 3C), correlation time (Fig 3D), and eigenvalue spectrum (Fig 3E). This is likely due to the heterogeneous length scales across neurons effectively smoothing out the transition in the homogeneous model. These results show how apparent scale-invariance can appear in a simple place cell model in the presence of heterogeneities.

\begin{figure*}[t]
\centering
\includegraphics[width=\textwidth]{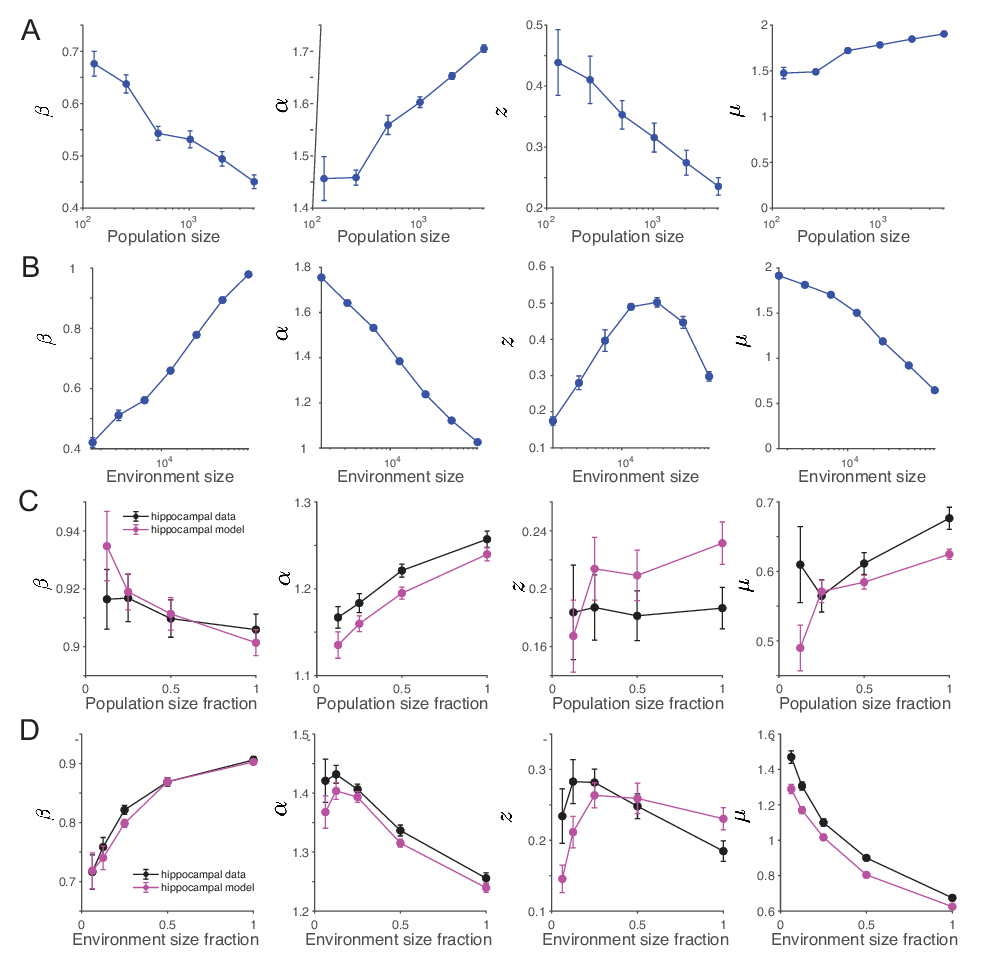}
\caption{Figure 4: Varying experimental parameters changes measured exponents. Changing the measured population size A) and environment size B) systematically changes the exponents measured from the simplified spatial coding model. Restricting population size C) and environment size D) affects the exponents measured in both hippocampal spatial coding model and hippocampal data.}
\end{figure*}

We next used this heterogeneous tile model to explore whether differences in the measured exponents between this and previous studies \cite{Meshulam2019} could be due to differences in the number of recorded neurons (fewer in this study) and/or environment size (larger in this study). When varying the population size, we observed that with a larger population size $\beta$ decreased, $\alpha$ and $z$ increased, and $\mu$ exhibited small changes (Figure 4A). When we varied environment size, we observed that increasing the environment size resulted in an increase of $\beta$, an inverted 'u' shaped curve for $z$, and a decrease of $\alpha$  and $\mu$ (Figure 4B). Hence, we speculate that differences in neural population and environment size could at least partially account for the differences between the exponents reported previously \cite{Meshulam2019} and those measured in our data (Fig 1I). 

The strong dependence of the power law exponents from the tile model on both environment and population size led us to reexamine our data and detailed place field model. We reasoned that a system with truly scale invariant properties should exhibit power law scaling and exponents that do not depend on these details. We hence sought to quantify the effect of changing these parameters on the exponents estimated from the data, by analyzing hippocampal responses from either smaller sections of the track or from a smaller number of recorded neurons. Decreasing population size (Figure 4C) increased $\beta$ and decreased $\alpha$, while reducing environment size (Figure 4D) increased $\alpha$ and $\mu$ while decreasing $\beta$, with an inverted 'u' shape for $z$. This is largely consistent with the predictions from the simple heterogeneous tile model (Figure 4A,B). Moreover, the dependency of all of the exponents on population and environment size were well captured by our detailed model (Figure 4 C,D, black curves). These results provide evidence against the hypothesis that power-law scaling observed in hippocampal data under PRG results directly from scale-invariant properties of hippocampal activity. 

\section{Discussion}
Our analysis of spatial coding models reveals that population heterogeneity alone can produce power law scaling under PRG with independent units. The simplified model we examined makes predictions about how the exponents depend on population size and environment size. We observed this dependency within our data set and noted that different population size and environment size explain differences between our experimentally measured exponents and other studies. Overall, this suggests that power-law scaling of hippocampal data under PRG does not constitute evidence for critical dynamics.

Previous models of place cells \cite{Meshulam2019, Morrell2021} relied on data taken in small environments where place cells typically exhibit 1 field with relatively small variation in place field size. We demonstrated analytically that this fails to produce power laws due to the emergence of two length scales: the environment size and the place field size. Their ratio, $N_{active}$, creates distinct scaling regimes for cluster sizes $K\ll N_{active}$ vs $K\gg N_{active}$. By contrast, our hippocampal model introduces a length scale to each neuron, interpreted as the average distance between fields. We showed that heterogeneities in the activity suffice to smooth over this kink, producing apparent power-law scaling. Although we used heterogeneity in place field size to achieve the smoothing in the tile model, our biological model used a more realistic heterogeneity in place field number\cite{Rich2014, Lee2020}, achieving exponents that more closely match experiment. 

We have noted above the similarity of the PRG to the renormalization procedure used to analyze coarsening phenomena \cite{Derrida1991}. It is however important to note an important difference between these two cases. For coarsening phenomena, the procedure reflect the intrinsic dynamic of the system, namely droplets merging and growing in size, as time evolves. On the contrary, for neural data, it is simply an exteriorly defined analysis procedure with no direct relation with neural dynamics. As such, the PRG has started to be applied to analyze neural data \cite{Ponce2023,Morales2023,Zivadinovic2025}. Our analysis has pointed the importance of place fields heterogeneity. Place cells and neuron tuning to space are of course prominent in the hippocampus but tuning of neurons to different quantities is widespread in the brain. Our analysis points out the general importance of analyzing the distribution of tuning curves, a more direct and intuitive measure of neural data than PRG exponents.

\section{End Matter}
\subsection{Heterogeneous place field model parameters}
\begin{table}[htbp]
    \centering
    \begin{tabular}{  c  c  c  c  c  c }
         Mouse & $a$ & $b$ (fields/environment)& $A$ (Hz) & $B$ (Hz) & $T$ (30 Hz bins) \\ \hline
         1 & .443 & .226 & 2.00 & .0020 & 11.0 \\
         2 & 2.37 & .969 & 2.75 & .0070 & 14.3 \\
         3 & 0.855 & .254 & 2.00 & .0030 & 11.7 \\
         4 & .805 & .575 & 2.25 & .010 & 11.1 \\
    \end{tabular}
    \caption{Heterogeneous model parameters fit for each animal. $a$ and $b$ are the shape and rate parameters, repectively, for the Gamma distribution describing heterogeneous field formation. $A$ and $B$ are the amplitudes of the spatially modulated and background calcium transient rates, respectively. $T$ is the exponential timescale describing each animal's observed transient durations.}
    \label{tab:parameters}
\end{table}

\subsection{Tile Model}
Initial unit activity for the homogeneous tile model is $\sigma_n^1(x)=\Theta\left( \frac{l}{2}-\left| x-\frac{L}{N}n \right|\right)$. It is useful to define $\delta x=L/N$, the spacing between unit activities, and use $K=2^k$ denoting the cluster size. Since the most correlated units are adjacent to one another, we can examine the statistics of a sum of adjacent units. The units grow with $K$ in amplitude until it reaches the number of active unit by location, $N_{active}$. While $1<K<N_{active}$, we have:
\begin{equation}
    P(\sigma_n^K=m)=
    \begin{cases} 
    1- \frac{l+(K-1)\delta x}{L} & m=0 \\
    \frac{2\delta x}{L} & 0 < m < K \\
    \frac{\delta x}{L} & m=K
    \end{cases}
\end{equation}

For $K\geq\rho$,
\begin{equation}
    P(\sigma_n^K=m)=
    \begin{cases} 
    1- \frac{l+(K-1)\delta x}{L} & m=0 \\
    \frac{2\delta x}{L} & 0 < m < N_{active} \\
    \frac{(2(K-N_{active})+1)\delta x}{L} & m=N_{active}
    \end{cases}
\end{equation}
The mean value $\langle\sigma_n^K\rangle_x=K\frac{l}{L}$. The probability of silence is therefore:
\begin{equation}
P_0=1-\frac{l+(K-1)\delta x}{L}
\end{equation}
This quantity shrinks exponentially with $k$, although an approximation of the 'power law' fit at any point can be estiated using $\beta_{est}=-\left. \frac{\partial \log P_0}{\partial \log K}\right|_{K=K^*}$. The variance can also be computed with different scaling relations for $K<N_{active}$ and $K>N_{active}$:
\begin{equation}
    Var_x(\sigma_n^K)=\begin{cases}
    \frac{\delta x}{3L}K(1-K^2)+\frac{l}{L}(1-\frac{l}{L})K^2 & K<N_{active} \\
    -\frac{l}{3L}(N_{active}^2-1)+\frac{l}{L}N_{active} K-\frac{l^2}{L^2}K^2 & K\geq N_{active}
    \end{cases}
\end{equation}
Once again, this is explicitly not a power law, although power law approximations can be made using $\alpha_{est}=\left.\frac{\partial \log Var_x(\sigma_n^K)}{\partial \log K}\right|_{K=K^*}$. Since these units are single-peaked (unlike real place cells), the autocorrelation function, $A^K(\Delta x)$, goes to negative values for large distances, rendering the exponential fit inapplicable. One instructive parameter regime to compute this is $K>\rho$, $N_{active}\delta x \leq \Delta x \leq (K-N_{active}+1)\delta x$, which gives:
\begin{equation}
    A^K(\Delta x)=N_{active}^2 \frac{(K+N_{active}-1)\delta x - \Delta x}{L}-\frac{l^2}{L^2}K^2
\end{equation}
This gives a linear fall-off, and describes the bulk of autocorrelation decay when $K\gg N_{active}$. The characteristic length of this process is $\tilde{x}=(K+N_{active}-1)\delta x$. A power law estimate can be made giving $z_{est}=\left. \frac{\partial \tilde{x}}{\partial \log K}\right|_{K=K^*} = \frac{K^*}{K^*+N_{active}-1}$. Another instructive example is $K<N_{active}$, $\Delta x<l(K-1)\delta x$. In this case, the autocorrelation function can be calculated:
\begin{equation}
    A^K(\Delta x)=\left[ l\left( 1-\frac{l}{L}\right)-\Delta x\right]\frac{K^2}{L}
\end{equation}
In this case, the characteristic length of the autocorrelation function is $\tilde{x}=(1-l/L)l$, and there is no $K$ dependence. We did not calculate the within-cluster eigenvalue spectrum for the tile model analytically.

\bibliography{mainbib}

\end{document}